%% file: preprint.tex
\documentclass{article} % For LaTeX2e
\usepackage{iclr2027_conference,times}

\IfFileExists{math_commands.tex}{\input{math_commands.tex}}{}
\usepackage{amsmath,amssymb}
\usepackage{amsthm}
\usepackage{stmaryrd}
\usepackage{booktabs}
\usepackage{graphicx}
\usepackage{adjustbox}
\usepackage{hyperref}
\usepackage{url}
\usepackage{wrapfig}
\usepackage{algorithmic}
\usepackage[table]{xcolor}
\usepackage{tabularx}
\usepackage{array}
\usepackage{fancyvrb}

\newcommand{\primarynum}[1]{{\normalsize\bfseries #1}}
\newcommand{\compnum}[1]{{\small #1}}

\definecolor{primaryblue}{RGB}{238,245,252}
\definecolor{resultblue}{RGB}{36,99,161}

\newtheorem{proposition}{Proposition}

\title{Purpose IR: Compiling User Purpose for Privacy-Preserving Agent Workflow State}

\author{
Dongsheng Chen$^{1}$ \quad
Xiangyu Zhao$^{2}$ \quad
Lei Ma$^{3}$ \quad
Xin Yao$^{4}$ \quad
Xuetao Wei$^{1}$\thanks{Corresponding author: \texttt{weixt@sustech.edu.cn}} \\
\\
$^{1}$Southern University of Science and Technology \\
$^{2}$City University of Hong Kong \\
$^{3}$The University of Tokyo \\
$^{4}$Lingnan University
}

\iclrfinalcopy

\begin{document}
\raggedbottom

\maketitle

% Override the ICLR camera-ready header for the arXiv preprint.
\fancyhead{}
\lhead{\textit{Preprint}}

\begin{abstract}
LLM-agent workflows routinely transform rich observations into plans,
summaries, memories, handoffs, and other reusable state. This creates a
privacy boundary that task relevance alone does not capture: an observation
may be needed as evidence for a task without its full semantics being
authorized to persist. We call this mismatch the
\emph{purpose-to-state semantic gap}. The key distinction is that
\emph{evidence authority is not release authority}. We introduce \textbf{Purpose IR}, a bounded typed intermediate representation
that compiles natural-language user purpose into an executable semantic
boundary for reusable agent state. A Purpose program is an acyclic semantic
dataflow graph in which evidence leaves specify task-required semantics,
typed dependencies constrain authorized derivations, and declared sinks
determine which values or actions may cross the reusable-state boundary.
A deterministic compiler constructs and validates the authoritative program,
while a typed runtime reconstructs reusable state only from validated sink
outputs. Under explicit execution contracts, this design yields
\emph{release confinement} and \emph{purpose noninterference},
characterizing what factual semantics may be released and which observation
differences must become irrelevant to reusable state. Across 1,200 cases spanning LangGraph, CrewAI, and MobileRun, Purpose IR
achieves \textbf{99.6\% Privacy}, \textbf{83.3\% Task}, and
\textbf{83.3\% Privacy-Compliant Success (PCS)}, outperforming the strongest
high-privacy baseline by \textbf{27.5 PCS points}. On 493 PrivacyLens
trajectories, it achieves the highest \textbf{Safe Action Rate} at
\textbf{77.3\%}, 10.4 points above the next-best method, extending the same
boundary from reusable state to downstream actions. These results show that
privacy-preserving agent state requires more than filtering sensitive or
task-irrelevant content: it requires an explicit, executable representation
of what the user's purpose authorizes to survive.
.
\end{abstract}

\section{Introduction}

LLM agents increasingly operate through multi-step workflows that combine language-model reasoning with tool use, environment observations, coordination, and persistent state \citep{yao2023react,schick2023toolformer,wu2024autogen,park2023generative,packer2023memgpt}. During execution, these workflows routinely transform rich observations into plans, summaries, memories, handoffs, and other intermediate artifacts that are reused by later steps or later agents. This reuse is operationally useful, but it creates a distinct privacy boundary: information that is legitimately observed for computation can persist far beyond the semantic result the user actually requested.

\begin{figure}[t]
\centering
\includegraphics[width=\linewidth]{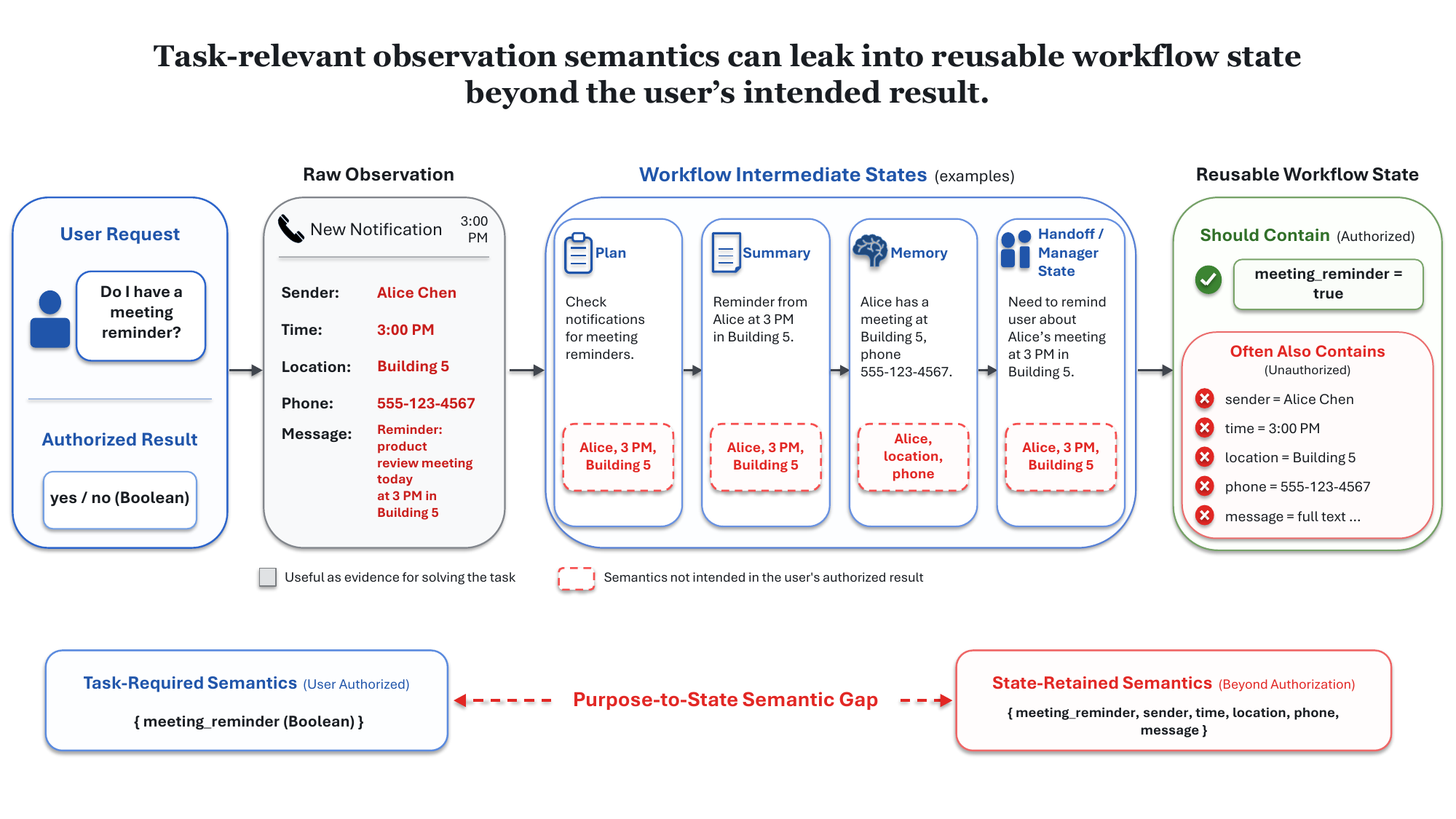}
\caption{\textbf{The purpose-to-state semantic gap.}
Task-relevant observation semantics may propagate through workflow intermediate states and persist in reusable state even when the user's authorized result is substantially narrower.}
\label{fig:purpose-state-gap}

\end{figure}

Consider a user who asks, ``Do I have a meeting reminder?'' A workflow may inspect a notification containing a sender name, meeting time, office location, phone number, and full message body. These fields can all help determine whether the notification is a meeting reminder, yet the requested result is only a Boolean predicate. As illustrated in Figure~\ref{fig:purpose-state-gap}, these task-relevant details can propagate through plans, summaries, memories, and handoff state, causing reusable state to preserve the semantics of the \emph{observation} rather than those of the \emph{requested result}. We call this mismatch the \textbf{purpose-to-state semantic gap}: the semantics retained in reusable workflow state can exceed the semantics required by the user's purpose. The key distinction is that the information an agent is authorized to inspect as evidence can be strictly richer than the information it is authorized to retain as reusable state. Task relevance therefore governs what may be useful as evidence, not what may persist as state.

The purpose-to-state semantic gap exposes three fundamental challenges.
\textbf{(P1) Semantic specification.}
A user's task specifies the desired outcome but not the boundary between semantics needed as evidence and those authorized to persist.
\textbf{(P2) Purpose compilation.}
Natural-language purpose is not directly executable and must be compiled into a representation that makes evidence, derivation, and release authority explicit.
\textbf{(P3) Semantic preservation.}
The compiled boundary must remain authoritative as information flows through heterogeneous workflow representations, preventing richer observation semantics from re-entering reusable state.
Existing span-based privacy tools, contextual-privacy models, runtime policies, and information-flow controls address important parts of these challenges, such as detecting sensitive content, judging contextual relevance, or restricting propagation \citep{microsoftPresidio,ngong2025contextualprivacy,bagdasarian2024airgap,costa2025fides,wang2026agentspec,siddiqui2025permissive,kolluri2026planning}. Existing studies also reveal leakage in reasoning traces, inter-agent communication, memories, and actions \citep{green2025leakythoughts,elyagoubi2026agentleak,shao2024privacylens}. What remains missing is a common, task-derived semantic boundary that specifies what result semantics may survive into reusable state and makes that boundary executable across workflows.

We introduce \textbf{Purpose IR} to provide this boundary. The core idea is simple: evidence may be richer than the result, but reusable state should not be. Purpose IR compiles a natural-language task into a bounded typed semantic dataflow program: evidence leaves specify what task semantics may be acquired, typed dependencies constrain how results may be derived, and declared sinks determine what values or actions may cross the reusable-state boundary. A task-only semantic frontend proposes candidate semantics, while a deterministic compiler constructs the authoritative program; workflow adapters ground its evidence requirements, and a typed runtime reconstructs reusable state only from validated sink outputs. Purpose IR therefore governs which semantics may survive, rather than how the agent performs the task.

This structure yields two formal properties that characterize the reusable-state boundary independently of any particular workflow representation. First, \emph{release confinement} requires every factual semantic value in reusable state to be justified by the authorized ancestry of an explicit sink. Second, \emph{purpose noninterference} requires observations that induce the same typed results at all sink-relevant Purpose leaves to produce equivalent reusable factual state under deterministic Purpose execution. Together, these properties capture both what information may be released and which observation differences must become irrelevant once the authorized result has been computed.

We evaluate Purpose IR in two complementary settings. The main evaluation contains 1,200 English workflow cases balanced equally across LangGraph, CrewAI, and MobileRun. Purpose IR achieves \textbf{99.6\% Privacy}, \textbf{83.3\% Task Success Rate}, and \textbf{83.3\% Privacy-Compliant Success (PCS)}; the strongest high-privacy baseline reaches 55.8\% PCS. We further evaluate external transfer on 493 PrivacyLens trajectories \citep{shao2024privacylens}, where Purpose IR achieves the highest \textbf{Safe Action Rate} at \textbf{77.3\%}, 10.4 points above the next-best method.

In summary, this paper makes three contributions aligned with the three challenges above:

\noindent\textbf{(1) A semantic formulation of reusable-state privacy.}
We identify the \textbf{purpose-to-state semantic gap} and distinguish task-relevant evidence from the smaller result semantics authorized to persist.\
\textbf{(2) Purpose compilation into an executable semantic representation.}
We introduce \textbf{Purpose IR}, a bounded typed dataflow IR that separates task-only semantic interpretation from deterministic authority construction and makes evidence, dependencies, and release/action sinks explicit.\
\textbf{(3) Semantic preservation with formal and cross-workflow evidence.}
We formalize the compiled reusable-state boundary through \emph{release confinement} and \emph{purpose noninterference}, and evaluate it across three workflow frameworks and on PrivacyLens.

\section{Related Work}

\paragraph{Privacy leakage in agent workflows.}
Recent work shows that privacy leakage in LLM agents is not confined to final outputs. \citet{green2025leakythoughts} demonstrate that reasoning traces of personal agents can contain sensitive user data, while AgentLeak instruments inter-agent messages, shared memory, tool arguments, and related channels, showing that output-only auditing can miss substantial internal leakage \citep{elyagoubi2026agentleak}. Multi-agent contextual-privacy work studies how upstream privacy errors propagate through workflow topology \citep{li2025onetwothree}, and CIPL models privacy leakage across heterogeneous agent pipelines as a channel-inversion problem \citep{huang2026cipl}. Together, these studies establish intermediate workflow artifacts and internal communication as important privacy surfaces. Purpose IR focuses on the semantics retained when legitimately accessed observations are transformed into reusable workflow state, and on how those retained semantics can be bounded by the user's task.

\paragraph{Task-aware privacy and agent enforcement.}
Several lines of work incorporate task context into privacy protection or enforce explicit constraints over agent execution. AirGapAgent minimizes the private source information exposed to a conversational agent \citep{bagdasarian2024airgap}, while contextual-privacy approaches evaluate disclosure in terms of relevance and necessity for the user's goal \citep{ngong2025contextualprivacy}. Fides and related work apply deterministic information-flow controls to agent planning and LLM calls \citep{costa2025fides,siddiqui2025permissive,kolluri2026planning}; AgentSpec provides a DSL for runtime safety constraints \citep{wang2026agentspec}; and agent firewalls, MNC, and CXI introduce structured or typed boundaries for agent communication and execution \citep{abdelnabi2025firewalls,xu2026mnc,santosgrueiro2026cxi}. AgentSecBench further separates data flow from authority and formalizes policy-enforced projections under intent-to-execution noninterference \citep{alpay2026agentsecbench}. Purpose IR contributes a task-derived semantic representation for this setting: natural-language purpose is compiled into typed evidence requirements, authorized derivations, and release sinks that govern the observation-to-state boundary.

\paragraph{Purpose, declassification, and policy languages.}
Purpose-based access control treats intended use as an explicit authorization dimension \citep{byun2008purpose,kebede2023purpose}. Classical information-flow control and declassification formalize restrictions on propagation and controlled release \citep{sabelfeld2003languagebased,sabelfeld2005declassification}, while systems such as Jeeves, Thoth, and Lifty use declarative policies or types to enforce privacy constraints \citep{yang2012jeeves,elnikety2016thoth,polikarpova2020lifty}. Purpose IR builds on this broader view of explicit release authority and instantiates it at the LLM-agent state boundary. In particular, it lowers natural-language user purpose into a bounded typed dataflow IR that specifies which evidence semantics may be acquired, how typed results may be derived, and which values or actions may be released into reusable state.

\section{Method}
\label{sec:method}

\begin{figure}[t]
    \centering
    \includegraphics[width=\linewidth]{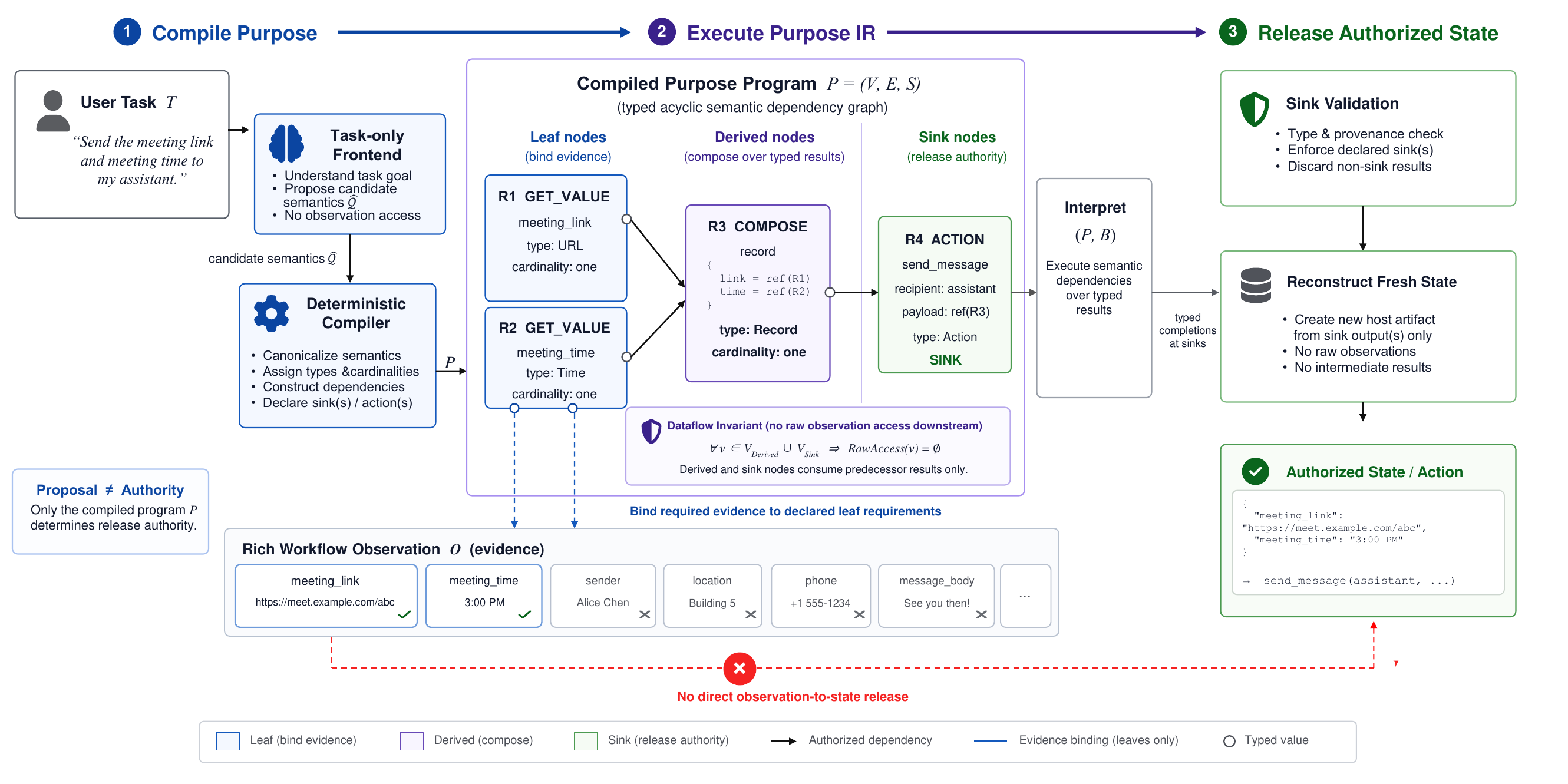}
    \caption{\textbf{Purpose IR overview.}
    User purpose is compiled into a typed semantic program that governs evidence binding, typed computation, and sink-constrained release.}
    \label{fig:purpose-ir-overview}
\end{figure}

Purpose IR makes the observation-to-state boundary executable. Given a user task \(T\) and workflow observation \(O\), the system compiles the task into a Purpose program, grounds the program in typed evidence, executes its semantic dataflow, and reconstructs only authorized outputs:
\[
P=\operatorname{Compile}(T),\qquad
B=\operatorname{Bind}(P,O),\qquad
C=\operatorname{Interpret}(P,B),\qquad
A_{\mathrm{out}}=\operatorname{Release}(P,C).
\]
The central separation is that \(O\) supplies evidence for the computation, whereas \(P\) defines the semantic authority under which results may enter reusable state. Figure~\ref{fig:purpose-ir-overview} gives an overview of this compiler--runtime design.

\subsection{Purpose IR: Typed Semantic Dataflow}
\label{sec:purpose-overview}

A compiled Purpose program is a typed acyclic graph
\[
P=(V,E,S),
\]
where \(V\) is a finite set of typed Purpose nodes, \(E\) is the authorized dependency relation, and \(S\) is the set of release or action sinks. Each node declares an operator, an output schema, and typed predecessor references; leaf nodes additionally declare a semantic target and evidence cardinality. Optional activation guards are bounded predicates over predecessor values. The graph captures \emph{semantic dependencies}, not agent control flow: tool selection, navigation, retry, and planning remain outside the IR.

Purpose nodes have three roles, summarized in Table~\ref{tab:purpose-node-roles}: evidence-binding leaves, typed derived nodes, and release/action sinks. Only leaves may bind workflow evidence; derived and sink nodes operate only on typed predecessor results.

\begin{table}[t]
\centering
\caption{Purpose IR node roles. Only leaf nodes may acquire workflow evidence; downstream nodes operate on typed predecessor outputs.}
\label{tab:purpose-node-roles}
\small
\setlength{\tabcolsep}{4pt}
\begin{adjustbox}{max width=\linewidth}
\begin{tabular}{llll}
\toprule
Role & Operators & Input authority & Output \\
\midrule
Leaf & \texttt{EXISTS}, \texttt{COUNT}, \texttt{GET\_VALUE}, \texttt{DIAGNOSE}
     & Declared workflow evidence & Typed result \\
Derived & \texttt{COMPOSE}
        & Typed predecessor results & Schema-bounded value \\
Sink & \texttt{ACTION} / direct release
     & Literals + typed predecessors & Reusable value or action \\
\bottomrule
\end{tabular}
\end{adjustbox}
\end{table}

This role separation creates the central dataflow invariant. For every derived or sink node \(v\),
\[
\operatorname{RawAccess}(v)=\varnothing,
\qquad
\operatorname{Inputs}(v)\subseteq
\{C_u:u\in\operatorname{Pred}(v)\}.
\]
Once observation content has been materialized as a typed Purpose result, downstream computation may use that result but cannot reopen the richer source observation that produced it.

For example, the task ``Send the meeting link and meeting time to my
assistant'' compiles to the following Purpose program:
\begin{Verbatim}[fontsize=\small,xleftmargin=1em]
R1 = GET_VALUE(meeting_link, one)
R2 = GET_VALUE(meeting_time, one)
R3 = COMPOSE(record,
             meeting_link=ref(R1),
             meeting_time=ref(R2))
R4 = ACTION(send_message,
            destination="assistant",
            content=ref(R3))
sinks = [R4]
\end{Verbatim}
The program exposes exactly the semantics authorized by the task---the meeting
link and time, their bounded composition, and a destination-scoped action---
without prescribing how the upstream agent discovers those values. The graph
representation becomes important when authorized results require multiple
semantic values to be composed, conditionally propagated, or bound to
structured actions. Purpose IR therefore represents the semantic computation
from task-required evidence to authorized release, rather than merely selecting
a flat set of fields.
\subsection{Compiling Natural-Language Purpose}
\label{sec:purpose-compilation}

Compilation separates semantic interpretation from executable release authority. A Qwen3.5-0.8B frontend reads only the user task \(T\) and proposes candidate operations, semantic targets, multiplicity, and decomposition:
\[
T\xrightarrow{M_{\theta}}\widehat{Q}
\xrightarrow{\operatorname{Compile}}P.
\]
The frontend proposal \(\widehat{Q}\) carries no execution or release authority. The deterministic Purpose Compiler resolves it into the canonical graph \(P\) by assigning semantic targets, operator contracts, schemas, cardinalities, dependencies, and sinks, and validating the result. It also separates task-level semantic targets, such as \texttt{meeting\_link} or \texttt{meeting\_time}, from generic observation carriers such as ``text'', ``notification'', or ``message body'', which cannot become release targets. Validation rejects malformed structure, including duplicate identifiers, unknown references, cycles, invalid guards or schemas, unsupported cardinalities, malformed action arguments, and derived references to raw observations. Once constructed, \(P\) is the sole authority consumed by the runtime; frontend proposals cannot directly expand evidence access or release.

The compiler provides executable authority rather than a guarantee that the resulting program perfectly captures the user's intended semantics. The formal properties below are therefore conditional on the compiled program \(P\) and the stated binding, operator, and reconstruction contracts; task-to-program semantic accuracy is evaluated empirically. Appendix~\ref{app:ir} and Appendix~\ref{app:compiler} give the complete node schema and lowering rules.

\subsection{Grounding and Typed Execution}
\label{sec:purpose-execution}

\paragraph{Requirement-directed evidence binding.}
Workflow representations differ, so Purpose IR separates semantic requirements from evidence carriers. For each leaf node \(v\), the workflow adapter produces transient bound evidence
\[
B_v=\operatorname{Bind}(v,O),
\]
annotated with semantic type, value, assertion status, source kind, and record identity.

\emph{Direct Binding} reads an explicitly declared native carrier when one is available. \emph{Recovery Binding} may recover a missing value from a sealed raw observation through a declared carrier mapping. Recovery changes where evidence is obtained, not what the Purpose program is authorized to materialize or release: the recovered value must still satisfy the node's semantic type, assertion constraints, cardinality, and codec. An adapter may inspect a declared carrier to recognize or extract the requested semantic value, but only the validated typed binding is materialized into Purpose execution state. For scalar targets, binding therefore returns the validated scalar rather than the larger carrier that contains it.

\paragraph{Topological typed interpretation.}
The interpreter executes nodes in topological order. For a node \(v\),
\[
C_v=
\llbracket v\rrbracket
\left(B_v,\{C_u:u\in\operatorname{Pred}(v)\}\right),
\]
with \(B_v=\varnothing\) for derived and sink nodes.

\texttt{EXISTS} distinguishes verified presence, verified absence, and unknown; \texttt{COUNT} operates over distinct matching record identities; \texttt{GET\_VALUE} validates type, codec, and declared cardinality; and \texttt{DIAGNOSE} produces only schema-declared bounded fields. \texttt{COMPOSE} can construct only values admitted by its output schema, and \texttt{ACTION} can consume only declared literals and typed predecessor references.

Ambiguous values, missing evidence, invalid cardinality, schema mismatch, or an incomplete sink leave the corresponding result unresolved rather than causing the runtime to widen the semantic release. Appendix~\ref{app:binding} and Appendix~\ref{app:semantics} specify the binding and operator contracts.

\subsection{Purpose-Constrained Release}
\label{sec:purpose-release}

Typed completions are computation state, not automatically reusable state. The release layer validates the declared sinks and reconstructs a fresh host-framework artifact from valid sink outputs only:
\[
A_{\mathrm{out}}=
\operatorname{Reconstruct}
\left(\{C_s:s\in S,\ C_s\text{ valid}\}\right).
\]
Source prose, evidence records, provenance, frontend candidates, compiler diagnostics, and non-sink completions do not become reusable merely because they participated in the computation.

Release is therefore a graph-level operation: factual content can enter reusable state only through validated sink outputs, and a typed dependency carries an authorized value without granting authority over unrelated fields from the same source observation. Dependency reachability and release authority are consequently distinct.

Returning to the meeting-reminder example, sender, time, location, phone number, and message body may all support the evidence-binding process for deciding whether a reminder exists, yet the reusable output can still be exactly the Boolean requested by the task. The richer carrier supports the computation without becoming part of the released result.

The Purpose language, interpreter, and release rule are shared across LangGraph, CrewAI, and MobileRun; workflow-specific code is confined to evidence adapters and native reconstruction. Appendix~\ref{app:adapters} and Appendix~\ref{app:formal} provide the implementation and formal details.

\subsection{Formal Properties}
\label{sec:formal-properties}

The graph discipline above yields two properties that characterize the semantic boundary enforced by Purpose IR. The properties concern factual semantics released into reusable state. Structured action sinks obey the same typed dependency and sink-validation discipline for their emitted arguments, while environment-level side effects are outside the scope of \(\operatorname{FactSem}\) and of the two propositions below. We call an execution \emph{valid} when leaf bindings respect their declared semantic targets and cardinalities, node evaluation respects the corresponding operator and schema contracts, and reconstruction emits factual content only from validated sink outputs.

\begin{proposition}[Release confinement]
\label{prop:release-confinement}
Let \(P=(V,E,S)\) be a valid Purpose IR program and consider a valid execution. For every factual semantic value \(x\) visible in the released artifact \(A_{\mathrm{out}}\), there exists a sink \(s\in S\) such that \(x\) is contained in the semantic output of \(s\) and its factual provenance is confined to semantics authorized by the Purpose ancestors of \(s\):
\[
x\in\operatorname{FactSem}(A_{\mathrm{out}})
\Longrightarrow
\exists s\in S:\;
x\in\operatorname{Sem}(C_s)
\;\land\;
\operatorname{Prov}(x)
\subseteq
\operatorname{Auth}(\operatorname{Anc}(s)).
\]
\end{proposition}

Release confinement states that graph reachability cannot create authority. A value may flow to a sink only through typed dependencies already admitted by the Purpose program; neighboring fields in the original observation do not become releasable merely because they appeared in the same evidence carrier or supported the same computation.

For the second property, let \(L_S\subseteq V\) denote the leaf ancestors of declared sinks. We call two observations \emph{purpose-equivalent} for \(P\) when they induce the same typed completion at every sink-relevant leaf:
\[
O_1\equiv_P O_2
\quad\Longleftrightarrow\quad
\forall \ell\in L_S,\;
C_\ell(O_1)=C_\ell(O_2).
\]

\begin{proposition}[Purpose noninterference]
\label{prop:purpose-noninterference}
For deterministic valid executions of \(P\), if \(O_1\equiv_P O_2\), then the two observations induce the same reusable factual semantic state:
\[
O_1\equiv_P O_2
\Longrightarrow
\operatorname{FactSem}(A_{\mathrm{out}}(O_1))
=
\operatorname{FactSem}(A_{\mathrm{out}}(O_2)).
\]
\end{proposition}

The proof follows the DAG structure: equal sink-relevant leaf completions imply equal derived completions by induction over topological order, hence equal sink outputs; deterministic sink-only reconstruction then yields equal factual semantic state. Intuitively, once two observations agree on everything the compiled purpose is authorized to use for its sinks, all remaining differences in the raw observations become irrelevant to reusable factual state. Appendix~\ref{app:formal} gives the execution contracts and proofs in detail.

\section{Experiments}
\label{sec:experiments}

\subsection{Evaluation Design}
\label{sec:exp-setup}

\paragraph{Main workflow-state benchmark.}
We evaluate on 1,200 English workflow cases spanning LangGraph, CrewAI, and
MobileRun, with 400 cases per workflow. The three subsets share a common
semantic task space while preserving their native workflow-state
representations. Cases cover semantic-boundary conditions such as ambiguous
evidence, negation, mixed private content, duplicate records, and semantic
minimality. For each case, all methods receive the same user task and the same
fixed workflow input.

We evaluate the downstream reusable artifact with a unified deterministic
evaluator. \textbf{Privacy} measures whether the artifact excludes
semantically unauthorized content, while \textbf{Task Success Rate} measures
whether task-required result semantics remain available downstream. Our
primary metric is their per-case conjunction, \textbf{Privacy-Compliant
Success (PCS)},
\[
\operatorname{PCS}(i)
=
\operatorname{Privacy}(i)
\land
\operatorname{TaskSuccess}(i).
\]

\paragraph{External transfer.}
We additionally evaluate on 493 PrivacyLens privacy-sensitive action
trajectories \citep{shao2024privacylens}. Unlike the main benchmark, this
setting evaluates downstream actions rather than reusable workflow state.
We follow the action-level protocol with a fixed local
Mistral-7B-Instruct-v0.2 judge, while Purpose IR uses the same task-only
Qwen3.5-0.8B semantic frontend as in the main evaluation.

We report \textbf{Privacy Pass}, \textbf{Helpfulness $\geq 2$}, and their
per-trajectory conjunction, \textbf{Safe Action Rate},
\[
\operatorname{SafeAction}(i)
=
\operatorname{PrivacyPass}(i)
\land
[\operatorname{Helpfulness}(i)\geq 2].
\]

\paragraph{Baselines.}
On the main benchmark, we compare against No Guard, Presidio
\citep{microsoftPresidio}, IBM Contextual Privacy LLM
\citep{ngong2025contextualprivacy}, Clean Task-text Guard,
Task-derived Structured Projection, Fides constrained inspection
\citep{costa2025fides}, and AgentSpec
\citep{wang2026agentspec}. On PrivacyLens, we compare against Full trajectory,
Task-text reduction, Presidio, AgentSpec, and Fides on the same 493
trajectories. Detailed baseline definitions, model roles, and evaluator
settings are provided in Appendix~\ref{app:baselines}.

\paragraph{Compositional compilation.}
Beyond end-to-end workflow performance, we evaluate purpose compilation on a
controlled 300-instance compositional suite requiring explicit semantic
dependencies, including argument binding, record composition, conditional
routing, grouping, and multi-stage dependency. We report validator-valid IR,
semantic-contract accuracy, and privacy, isolating whether the frontend and
compiler recover the intended computation rather than merely the authorized
leaf semantics.

\subsection{End-to-End Privacy--Utility Results}
\label{sec:end-to-end-results}

Table~\ref{tab:end-to-end-results}(a) reports results on the main
workflow-state benchmark. Purpose IR reaches \textbf{99.6\% Privacy},
\textbf{83.3\% Task Success Rate}, and \textbf{83.3\% Privacy-Compliant Success (PCS)},
improving PCS by \textbf{27.5 percentage points} over Fides at a comparable
privacy level. The baselines expose the expected privacy--utility trade-off:
broad retention preserves task information but also unauthorized semantics,
whereas conservative filtering suppresses task utility. Structured Projection
reaches only 24.4\% PCS, showing that task-aware field selection alone is
insufficient. Across individual workflows, Privacy ranges from 99.5--99.8\%
and PCS from 79.8--85.5\%, with 100\% validator-valid IR across all three
frameworks; see Table~\ref{tab:per-workflow-results} in the appendix. Figure~\ref{fig:privacy-utility-frontiers}(a) visualizes this
high-privacy, high-utility operating point; the Purpose language, compiler, and
interpreter are shared across workflows, with only evidence grounding and
native reconstruction being adapter-specific.

\paragraph{External transfer.}
Table~\ref{tab:end-to-end-results}(b) and
Figure~\ref{fig:privacy-utility-frontiers}(b) show the same pattern on
493 PrivacyLens trajectories. Purpose IR achieves the highest
\textbf{Safe Action Rate} at \textbf{77.3\%}, \textbf{10.4 percentage points}
above Task-text reduction. Full trajectory favors utility and Fides favors
privacy, whereas Purpose IR achieves the strongest conjunction of the two.
Because PrivacyLens changes both the input representation and the evaluated
release object, this result provides external evidence beyond the main
workflow-state benchmark.

\begin{figure}[t]
    \centering
    \begin{minipage}[t]{0.49\linewidth}
        \centering
        \textbf{(a) Main benchmark}\\[-2pt]
        \includegraphics[width=\linewidth]{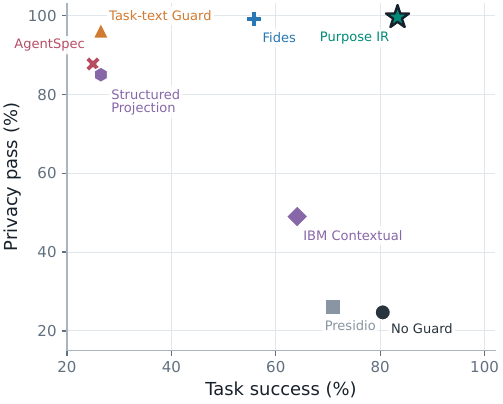}
    \end{minipage}
    \hfill
    \begin{minipage}[t]{0.49\linewidth}
        \centering
        \textbf{(b) PrivacyLens}\\[-2pt]
        \includegraphics[width=\linewidth]{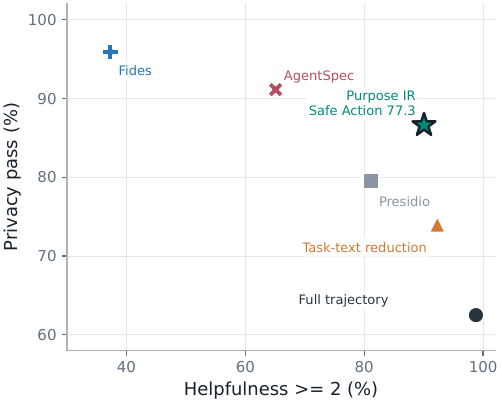}
    \end{minipage}
    \caption{\textbf{Privacy--utility trade-offs.}
    (a) Task Success Rate versus Privacy on the main benchmark.
    (b) Helpfulness versus Privacy Pass on PrivacyLens.}
    \label{fig:privacy-utility-frontiers}
    \vspace{-4pt}

\end{figure}

\begin{table}[t]
\centering
\caption{\textbf{End-to-end results.}
PCS and Safe Action require privacy and utility on the same case or trajectory.}
\label{tab:end-to-end-results}

\small
\setlength{\tabcolsep}{4pt}
\renewcommand{\arraystretch}{1.08}

% ==================== Left panel ====================
\begin{minipage}[t]{0.485\linewidth}
\vspace{0pt}
\centering

\textbf{(a) Main benchmark}\\[-2pt]

\begin{tabularx}{\linewidth}
{@{}>{\raggedright\arraybackslash}Xr@{\hspace{8pt}}rr@{}}
\toprule
&
\multicolumn{1}{c}{\cellcolor{primaryblue}\textbf{Primary}}
&
\multicolumn{2}{c}{Secondary} \\
\cmidrule(lr){2-2}
\cmidrule(lr){3-4}
Method
& \cellcolor{primaryblue}\textbf{PCS $\uparrow$}
& Privacy
& Task Success Rate \\
\midrule
No Guard
& \primarynum{23.9} & \compnum{24.7} & \compnum{80.5} \\

Presidio
& \primarynum{20.8} & \compnum{26.0} & \compnum{71.0} \\

IBM Context.
& \primarynum{32.3} & \compnum{49.0} & \compnum{64.1} \\

Task-text
& \primarynum{25.3} & \compnum{96.2} & \compnum{26.5} \\

Struct. Proj.
& \primarynum{24.4} & \compnum{85.0} & \compnum{26.5} \\

Fides
& \primarynum{55.8} & \compnum{99.1} & \compnum{55.9} \\

AgentSpec
& \primarynum{16.2} & \compnum{87.7} & \compnum{25.0} \\

\midrule
\textbf{Purpose IR}
& {\color{resultblue}\primarynum{83.3}}
& \compnum{99.6}
& \compnum{83.3} \\
\bottomrule
\end{tabularx}

\end{minipage}
\hfill
% ==================== Right panel ====================
\begin{minipage}[t]{0.485\linewidth}
\vspace{0pt}
\centering

\textbf{(b) PrivacyLens}\\[-2pt]

\begin{tabularx}{\linewidth}
{@{}>{\raggedright\arraybackslash}Xr@{\hspace{8pt}}rr@{}}
\toprule
&
\multicolumn{1}{c}{\cellcolor{primaryblue}\textbf{Primary}}
&
\multicolumn{2}{c}{Secondary} \\
\cmidrule(lr){2-2}
\cmidrule(lr){3-4}
Method
& \cellcolor{primaryblue}\textbf{Safe Action $\uparrow$}
& Priv. Pass
& Help.$\geq$2 \\
\midrule
Full traj.
& \primarynum{61.7} & \compnum{62.5} & \compnum{98.8} \\

Task-text
& \primarynum{66.9} & \compnum{74.0} & \compnum{92.3} \\

Presidio
& \primarynum{62.3} & \compnum{79.5} & \compnum{81.1} \\

AgentSpec
& \primarynum{56.2} & \compnum{91.1} & \compnum{65.1} \\

Fides
& \primarynum{34.1} & \compnum{95.9} & \compnum{37.3} \\

\midrule
\textbf{Purpose IR}
& {\color{resultblue}\primarynum{77.3}}
& \compnum{86.6}
& \compnum{90.1} \\
\bottomrule
\end{tabularx}

\end{minipage}

\end{table}

\subsection{Mechanism Analysis}
\label{sec:mechanism-results}

\begin{figure}[t]
    \centering
    \includegraphics[width=\linewidth]{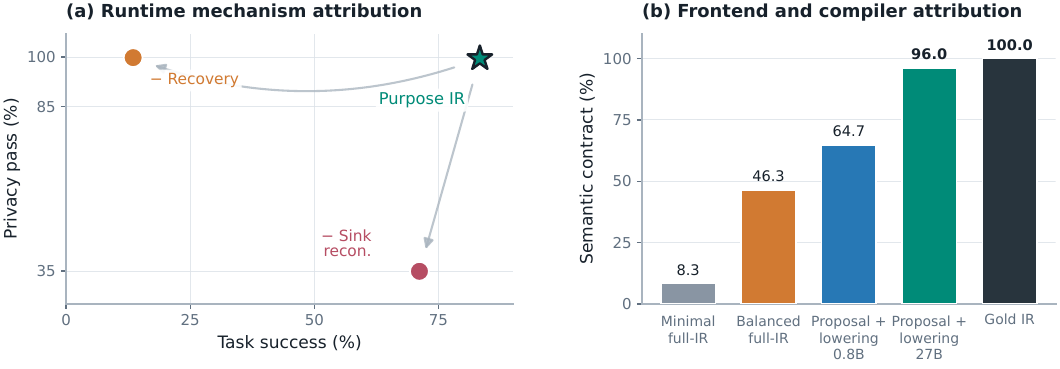}
    \caption{\textbf{Mechanism and compilation attribution.}
        (a) Paired runtime ablations on the 1,200-case main benchmark.
        (b) Semantic-contract accuracy on the controlled 300-instance
        compositional suite.}
    \label{fig:mechanism-map}
\end{figure}

We isolate two complementary sources of performance: runtime enforcement on
the main workflow benchmark and semantic compilation on the compositional
suite. Figure~\ref{fig:mechanism-map} summarizes both analyses.

\paragraph{Runtime mechanisms.}
All runtime variants replay the same 1,200 compiled Purpose programs with no additional
model calls and use the same deterministic evaluator as Full Purpose IR.
Removing recovery binding leaves Privacy at 99.8\% but reduces Task Success
Rate and PCS from 83.3\% to \textbf{13.5\%}. In contrast, removing
sink-constrained reconstruction retains 71.2\% Task Success Rate but reduces
Privacy from 99.6\% to \textbf{34.9\%} and PCS to \textbf{24.2\%}. The mean
released artifact grows from 203.16 to 679.40 bytes without sink-constrained
reconstruction. These paired interventions separate the two runtime roles:
recovery binding supplies task-required typed evidence, whereas reconstruction
determines which computed semantics may persist.

\paragraph{Frontend and compiler.}
On the controlled compositional suite, semantic-contract accuracy rises from
25/300 with minimal direct full-IR prompting to 139/300 with balanced
prompting, and to \textbf{194/300} with 0.8B semantic proposal plus
deterministic lowering. Under the same compilation and evaluation protocol, a
27B semantic frontend reaches \textbf{288/300}. Privacy is 300/300 for all evaluated variants. This progression
separates executable IR construction from semantic interpretation and shows
the benefit of deterministic lowering together with stronger semantic
frontends.

Finally, we validate the language boundary independently of end-to-end task
performance. Contract-mutation tests reject invalid cardinalities, dangling
dependencies, raw-observation references, invalid sinks, and dependency
cycles before execution. Cardinality-specific tests further confirm that
ambiguous \texttt{cardinality=one} evidence is withheld rather than widened
into a multi-value release. Together, these checks verify that the structural
constraints assumed by Purpose IR are enforced by the implementation.
Detailed intervention results are reported in
Appendix~\ref{app:mechanism-details}.

\section{Conclusion}

We introduced \textbf{Purpose IR}, a compiler-based approach to governing
what semantics may persist in reusable LLM-agent state. Its central distinction
is between evidence authority and release authority: rich observations may be
needed to compute a task result without their full semantics being authorized
to persist. Purpose IR makes this boundary executable through typed semantic
dependencies and explicit release sinks, with \emph{release confinement} and
\emph{purpose noninterference} characterizing the resulting reusable-state
semantics. Our evaluations show that this boundary can preserve task utility
while substantially reducing unauthorized retention. More broadly, agent
privacy should govern not only what information may be accessed, but also what
semantics the user's purpose authorizes to survive.

\subsection*{AI Use Statement}

We partially used generative AI tools for code development and for improving the clarity and readability of the manuscript. All AI-assisted code and text modifications were carefully reviewed, verified, and tested by the authors.

\section*{Ethics statement}

This work does not involve human subjects, sensitive personal data, or applications that raise
significant ethical concerns. We have considered potential ethical risks associated with this research
and found no additional issues requiring further discussion.

\section*{Reproducibility statement}

We are committed to making this work reproducible. In the main paper and appendix, we provide as much
detail as possible about the method, model architecture, and experimental details needed to reproduce
this work, and we provide an anonymized code repository link to facilitate reproduction.
\bibliography{iclr2027_conference}
\bibliographystyle{iclr2027_conference}

\appendix

\section{Purpose IR Specification}
\label{app:ir}

This appendix specifies the bounded typed dataflow language used by Purpose IR and the static constraints enforced before execution.

\subsection{Language boundary}
\label{app:ir-boundary}

Purpose IR is a semantic dataflow language for user purpose, not a full agent execution language. It represents task-required evidence semantics, typed dependencies among authorized intermediate results, and declared release or action sinks. Tool selection, navigation strategy, retry policy, loops, recursion, and unrestricted user-defined computation remain outside the IR.

A Purpose program is
\[
P=(V,E,S),
\]
where \(V\) is a finite set of typed nodes, \(E\) is an acyclic dependency relation, and \(S\) is the set of declared sinks. The operator vocabulary is partitioned by role:
\[
\mathcal{O}_{\mathrm{leaf}}=
\{\texttt{EXISTS},\texttt{COUNT},\texttt{GET\_VALUE},\texttt{DIAGNOSE}\},
\]
\[
\mathcal{O}_{\mathrm{derived}}=\{\texttt{COMPOSE}\},
\qquad
\mathcal{O}_{\mathrm{sink}}=\{\texttt{ACTION}\}.
\]
A completed leaf or derived node may also be declared directly as a release sink when the task result itself is the reusable output.

\subsection{Canonical node contract}
\label{app:canonical-contract}

Each node has a stable identifier, operator, output schema, and optional dependency metadata. Leaf nodes additionally declare a semantic target and evidence cardinality. Derived and sink nodes reference predecessor results explicitly. Table~\ref{tab:req-schema} summarizes the core fields.

A representative compositional program is:
\begin{verbatim}
R1 = GET_VALUE(error_code, cardinality=one)
R2 = GET_VALUE(device_model, cardinality=one)
R3 = COMPOSE(record,
             error_code=ref(R1),
             device_model=ref(R2))
R4 = ACTION(send_message,
            destination="administrator",
            content=ref(R3))
sinks = [R4]
\end{verbatim}
The graph makes internal and external values explicit. \(R1\)--\(R3\) can complete without becoming reusable state; \(R4\) is the declared sink.

\subsection{Semantic targets and cardinality}
\label{app:targets}

Semantic targets belong to a bounded registered namespace with auditable binding paths. Examples include \texttt{meeting\_reminder}, \texttt{meeting\_link}, \texttt{tracking\_number}, \texttt{error\_code}, \texttt{device\_model}, \texttt{app\_version}, and \texttt{delivery\_status}. Generic carriers such as \texttt{text}, \texttt{clipboard}, \texttt{notification}, \texttt{screen}, or \texttt{message\_body} are not semantic targets.

Leaf nodes additionally declare bounded evidence cardinality. The implementation supports single-value and bounded multi-value requirements, including \texttt{one}, \texttt{many}, \texttt{at\_most\_k}, and \texttt{all\_matching}. Cardinality is checked during binding and materialization rather than inferred from field order at runtime.

\subsection{Dependency, composition, and sink semantics}
\label{app:composition}

Dependencies are first-class typed references. A node may consume only declared predecessor outputs, and the compiler rejects unknown references and cycles. Derived \texttt{COMPOSE} nodes construct schema-bounded records or templates from predecessor values; they do not receive an observation handle. Structured \texttt{ACTION} nodes similarly consume declared literals and typed predecessor results and define the downstream destination and arguments.

This separation yields the leaf-access invariant
\[
v\in V_{\mathrm{derived}}\cup V_{\mathrm{sink}}
\Longrightarrow
\operatorname{RawAccess}(v)=\varnothing.
\]
A dependency therefore transports an already-authorized typed value; it does not expand the semantic authority of the consumer.

\section{Operator and Failure Semantics}
\label{app:semantics}

Table~\ref{tab:operator-contracts} summarizes the typed operator roles.

\begin{table}[t]
\centering
\caption{Typed Purpose IR operator contracts.}
\label{tab:operator-contracts}
\small
\begin{adjustbox}{max width=\linewidth}
\begin{tabular}{llll}
\toprule
Operator & Role & Input & Output \\
\midrule
\texttt{EXISTS} & Leaf & Verified predicate evidence & Boolean \\
\texttt{COUNT} & Leaf & Verified matching records & Non-negative integer \\
\texttt{GET\_VALUE} & Leaf & Validated typed evidence & Typed scalar/collection \\
\texttt{DIAGNOSE} & Leaf & Valid schema-matched evidence & Bounded structured result \\
\texttt{COMPOSE} & Derived & Typed predecessor outputs & Schema-bounded value \\
\texttt{ACTION} & Sink & Literals + typed predecessor outputs & Structured action arguments \\
\bottomrule
\end{tabular}
\end{adjustbox}
\end{table}

\subsection{Three-valued existence semantics}
\label{app:exists}

The runtime distinguishes verified absence from insufficient evidence:
\[
s\in\{\mathrm{PRESENT},\mathrm{ABSENT},\mathrm{UNKNOWN}\}.
\]
The operator semantics are
\[
\operatorname{EXISTS}(s)=
\begin{cases}
\mathrm{true}, & s=\mathrm{PRESENT},\\
\mathrm{false}, & s=\mathrm{ABSENT},\\
\bot, & s=\mathrm{UNKNOWN}.
\end{cases}
\]
A verified negative answer is a valid task result. Missing or conflicting evidence remains unresolved.

\subsection{Counting, values, and cardinality}
\label{app:count}

\texttt{COUNT} accepts a validated non-negative aggregate when the adapter establishes that it represents the requested count; otherwise it counts distinct matching records. With stable record identities,
\[
\operatorname{COUNT}(E)=
\left|\{\operatorname{record\_id}(e):e\in E\}\right|.
\]
\texttt{GET\_VALUE} validates semantic type, codec, and declared cardinality. A \texttt{one} requirement must resolve to one unique value; bounded multi-value modes validate their declared limits. Conflicting candidates fail closed rather than being resolved by iteration order.

A scalar result is released as the validated scalar, not as a larger carrier containing adjacent semantics. For example,
\begin{verbatim}
value=E42; workspace=group-alpha; owner=member-7
\end{verbatim}
is not a valid scalar representation of \texttt{E42}.

\subsection{Bounded diagnosis and composition}
\label{app:diagnose}

\texttt{DIAGNOSE} produces only fields declared by its output schema. It is not a license to copy raw debug prose, screenshots, source quotes, or complete native records. \texttt{COMPOSE} operates only over predecessor completions and must satisfy its declared result schema. The derived node has no raw-observation interface.

\subsection{Structured action sinks}
\label{app:action}

\texttt{ACTION} is a sink rather than an unrestricted text generator. Its destination and argument schema are declared by the Purpose program, and each argument is either a contract literal or a typed reference to a predecessor completion. The runtime validates the complete sink before release. A sink cannot widen its inputs by consulting source text, raw observations, or undeclared intermediate state.

\begin{table}[t]
\centering
\caption{Canonical Purpose IR node fields.}
\label{tab:req-schema}
\small
\begin{adjustbox}{max width=\linewidth}
\begin{tabular}{lll}
\toprule
Field & Meaning & Deterministic constraint \\
\midrule
\texttt{id} & Stable node identifier & Unique within the graph \\
\texttt{operator} & Purpose operation & Registered operator \\
\texttt{target\_type} & Semantic evidence type for leaf nodes & Registered semantic target \\
\texttt{output\_schema} & Typed result shape & Compatible with operator \\
\texttt{cardinality} & Evidence multiplicity & Declared bounded mode \\
\texttt{depends\_on} & Predecessor node identifiers & Existing nodes; acyclic graph \\
\texttt{condition} & Optional typed activation guard & References typed predecessor results \\
\texttt{arguments} & Structured inputs for composition/action & Literals or typed references only \\
\texttt{release} & Node-local release metadata & Consistent with sink declaration \\
\texttt{grounding} & Evidence-binding metadata & Descriptive; grants no authority \\
\bottomrule
\end{tabular}
\end{adjustbox}
\end{table}
\subsection{Representative codecs}
\label{app:codecs}

\begin{table}[t]
\centering
\caption{Representative result codecs.}
\label{tab:codecs}
\small
\begin{adjustbox}{max width=\linewidth}
\begin{tabular}{lll}
\toprule
Codec & Accepted output & Rejected output \\
\midrule
\texttt{boolean} & Boolean value & Arbitrary prose or record-shaped text \\
\texttt{integer} & Non-negative integer & Negative or malformed number \\
\texttt{uri} & Exactly one HTTP(S) URI & Zero/multiple URIs or mixed span \\
\texttt{bounded\_scalar} & Single typed scalar & Dict/list/multiline carrier \\
\texttt{object/array} & Schema-valid structured value & Extra or ill-typed fields \\
\bottomrule
\end{tabular}
\end{adjustbox}
\end{table}

\subsection{Fail-closed behavior}
\label{app:failure}

The runtime treats uncertainty or invalid graph structure as a reason to withhold release rather than widen it. Representative failures include unsupported operators or targets, duplicate identifiers, unknown dependencies, dependency cycles, invalid conditions, unsupported cardinality, ambiguous evidence, codec failure, schema mismatch, missing action arguments, derived raw-observation references, incomplete sinks, and reconstruction postcondition violations. Graph-level validation therefore extends the value-level fail-closed behavior of the leaf operators to the entire Purpose program.

\section{Semantic Frontend and Purpose Compiler}
\label{app:compiler}

The compiler separates neural semantic proposal from deterministic authority construction.

\subsection{Task-only semantic frontend}
\label{app:frontend}

The Qwen3.5-0.8B frontend reads only the natural-language task. It does not observe native workflow state, raw observations, artifact fields, privacy labels, or benchmark annotations. It proposes candidate operations, semantic targets, multiplicity, and task decomposition for the compiler. The proposal carries no release authority.

\subsection{Deterministic lowering and static validation}
\label{app:lowering}

The Purpose Compiler normalizes operator and target names, validates targets against the semantic registry, constructs leaf requirements, assigns output schemas and cardinalities, resolves dependencies, constructs structured sinks, and checks the resulting DAG. Static validation rejects duplicate identifiers, unknown references, cycles, incompatible schemas, invalid cardinality, malformed typed guards, illegal sink arguments, and derived references to raw observations.

Only the compiler-admitted canonical graph is consumed by the runtime.

\subsection{Scope of the language}
\label{app:compiler-scope}

Purpose IR is deliberately bounded around semantic computation and release. It supports typed evidence requirements, acyclic semantic dependencies, bounded cardinality, typed activation conditions, schema-bounded composition, and structured sinks. General agent control flow---including loops, recursion, arbitrary tool planning, unrestricted expressions, and mutable program state---remains the responsibility of the host agent rather than the Purpose IR.

\section{Fact IR, Binding, and Recovery}
\label{app:binding}

Purpose IR is workflow-independent, but evidence representations are not. The
runtime therefore converts native workflow state into a common transient
evidence interface before typed execution.

\subsection{Fact IR}
\label{app:fact-ir}

The native draft is atomized into complete information units. A representative
Fact IR atom is:
\begin{verbatim}
{
  "atom_id": "a17",
  "fact_id": "f17",
  "source_path": ["answer", "text"],
  "role": "CONTENT",
  "assertion_status": "ASSERTED",
  "structural_context": "answer.text",
  "proposition":
      "meeting_link=https://meet.example.com/j/8421"
}
\end{verbatim}

Fact IR separates semantic content from processing metadata. Source paths,
provenance, source quotes, control schema, and reasoning traces remain internal
metadata; their presence does not make them factual task outputs.

\subsection{Bound evidence}
\label{app:bound-evidence}

The shared interpreter consumes transient \texttt{BoundEvidence} objects:
\begin{verbatim}
{
  "requirement_id": "R1",
  "semantic_type": "meeting_link",
  "value": "https://meet.example.com/j/8421",
  "source_kind": "direct_binding",
  "assertion_status": "ASSERTED",
  "role": "CONTENT",
  "record_id": "record-17"
}
\end{verbatim}

Before materialization, the runtime validates the requirement identifier,
assertion state, evidence role, and semantic type. Bound evidence is transient;
it is not itself reusable workflow state.

\subsection{Direct and recovery binding}
\label{app:direct-recovery}

\paragraph{Direct Binding.}
Direct Binding obtains typed evidence from structured or already exposed
workflow state. It is preferred when the adapter has an explicit carrier for
the requested semantic type.

\paragraph{Recovery Binding.}
When Direct Binding is insufficient, the adapter may recover missing evidence
from the sealed raw observation. Recovery is requirement-directed and uses
declared carrier mappings. A recovered value must still match the requirement
identifier and semantic type, have an admissible role and assertion state, and
pass the corresponding codec.

Recovery therefore changes where evidence is obtained, not what the task is
authorized to release.

\subsection{Representative carrier mappings}
\label{app:carrier-map}

\begin{table}[t]
\centering
\caption{Representative semantic targets and adapter-owned evidence carriers.}
\label{tab:carrier-map}
\small
\begin{adjustbox}{max width=\linewidth}
\begin{tabular}{llll}
\toprule
Semantic target & Representative carrier & Codec & Validation \\
\midrule
\texttt{meeting\_link} & Declared URL field/carrier & \texttt{uri} & Exactly one URI \\
\texttt{tracking\_number} & Declared value field & \texttt{bounded\_scalar} & Single-line scalar \\
\texttt{error\_code} & Error-code carrier & \texttt{bounded\_scalar} & No mixed record span \\
\texttt{device\_model} & Device/debug record field & \texttt{bounded\_scalar} & Exact semantic type \\
\texttt{app\_version} & Version field & \texttt{bounded\_scalar} & Single-line scalar \\
\texttt{meeting\_reminder} & Typed notification category & \texttt{boolean} & Predicate + assertion \\
\texttt{unread\_count} & Count/unread aggregate & \texttt{integer} & Non-negative integer \\
\texttt{crash\_stack} & Declared debug field & \texttt{structured\_summary} & Structured evidence \\
\bottomrule
\end{tabular}
\end{adjustbox}
\end{table}

Carrier aliases belong to workflow adapters. The shared Purpose interpreter does
not contain native field names.

\subsection{Ambiguity and mixed spans}
\label{app:mixed-span}

For scalar and action requirements, adapters may rank declared candidate
carriers. If two equally admissible candidates contain conflicting values, the
runtime returns no binding instead of choosing one according to iteration or
field order.

Likewise, a prose span is not accepted as a scalar merely because the desired
value appears within it. URI extraction requires one unambiguous URI, and the
bounded scalar codec rejects multiline or record-shaped spans containing
adjacent \texttt{field=value} material. This constraint is semantic rather than
a blacklist of private keywords: the released object must have the shape
authorized by the Purpose requirement.

\section{Workflow Adapters}
\label{app:adapters}

The Purpose language and interpreter are shared across workflows, while
representation-specific operations remain in adapters. Table~\ref{tab:shared}
summarizes this boundary.

\begin{table}[t]
\centering
\caption{Shared and workflow-specific components.}
\label{tab:shared}
\small
\begin{adjustbox}{max width=\linewidth}
\begin{tabular}{lcc}
\toprule
Component & Shared & Workflow-specific \\
\midrule
Purpose language & \checkmark & \\
Operator semantics & \checkmark & \\
Compiler contract validation & \checkmark & \\
Typed codecs & \checkmark & \\
Release-authority checks & \checkmark & \\
Fact IR atomization & & \checkmark \\
Native carrier aliases & & \checkmark \\
Direct Binding & & \checkmark \\
Recovery Binding & & \checkmark \\
Native reconstruction & & \checkmark \\
Workflow contamination checks & & \checkmark \\
\bottomrule
\end{tabular}
\end{adjustbox}
\end{table}

\paragraph{LangGraph.}
LangGraph cases expose structured graph-state surfaces. The adapter traverses
native state, atomizes content with path information, maps declared typed
carriers into BoundEvidence, and reconstructs graph-compatible state from
authorized results.

\paragraph{CrewAI.}
CrewAI cases expose task, description, result, plan, answer, and handoff-like
prose surfaces. The adapter treats them as separate native carriers during
atomization and reconstruction. Repeated prose surfaces do not survive merely
because they are part of the native task object.

\paragraph{MobileRun.}
MobileRun cases expose manager, executor, memory, action, and related state
surfaces. The adapter maps these native paths into the same evidence interface
and reconstructs only the authorized result into the host schema.

The language, operator semantics, and release rule remain fixed across the three workflows; adapters provide the representation-specific translation into and out of the shared typed evidence interface. This separation is the portability boundary evaluated in our main suite.

\section{Formal Properties}
\label{app:formal}

This section states the execution contracts and proof obligations behind Propositions~\ref{prop:release-confinement} and~\ref{prop:purpose-noninterference}. These contracts define valid execution for the shared Purpose IR semantics.

\subsection{Execution Contracts}
\label{app:formal-assumptions}

Let $P=(V,E,S)=\operatorname{Compile}(T)$ be a valid acyclic Purpose program and $O$ a workflow observation. A valid execution satisfies the following contracts.
\begin{enumerate}
    \item \textbf{Binding contract.} For every leaf $\ell$, each value in $B_\ell=\operatorname{Bind}(\ell,O)$ denotes the semantic target declared by $\ell$, satisfies its assertion constraints, and preserves record identity when required by the operator.
    \item \textbf{Operator and codec contract.} For every node $v$, $\llbracket v\rrbracket$ produces only values admitted by the node's declared output type, cardinality, and schema. Derived and sink nodes read only typed predecessor outputs and declared literals.
    \item \textbf{Reconstruction contract.} $\operatorname{Reconstruct}$ emits factual semantic content only from validated sink outputs; source observations, bound evidence, provenance, compiler diagnostics, and non-sink completions are not copied into factual reusable state.
    \item \textbf{Determinism} (for Proposition~\ref{prop:purpose-noninterference}). Given the same typed predecessor completions, guards, operator semantics, and sink outputs, graph execution and factual reconstruction are deterministic.
\end{enumerate}

Together, these contracts define the executions to which the two propositions apply.

\subsection{Proof of release confinement}
\label{app:proof-release}

For each non-leaf node $v$, the language invariant requires
\[
\operatorname{RawAccess}(v)=\varnothing,
\qquad
\operatorname{Inputs}(v)\subseteq\{C_u:u\in\operatorname{Pred}(v)\}.
\]
Consider any $x\in\operatorname{FactSem}(A_{\mathrm{out}})$. By the reconstruction contract, $x$ must be emitted from a validated sink output $C_s$ for some $s\in S$. Each factual component of $C_s$ is either a contract literal or is obtained through a typed predecessor reference. Following predecessor references backward in the finite acyclic graph reaches only leaf completions and declared literals. By the binding contract, each leaf completion denotes its declared semantic target; by the operator and codec contract, derived nodes cannot introduce factual inputs outside their declared predecessor semantics. Therefore every factual provenance path for $x$ is contained in semantics authorized by $\operatorname{Anc}(s)$. This establishes Proposition~\ref{prop:release-confinement}.

The result is a confinement property of the language and release discipline: dependency edges transport already admitted values, but do not grant authority over other semantics co-located in the same observation carrier.

\subsection{Proof of purpose noninterference}
\label{app:proof-noninterference}

Let $L_S$ be the leaf ancestors of the declared sinks and suppose $O_1\equiv_P O_2$. By definition,
\[
\forall \ell\in L_S,\qquad C_\ell(O_1)=C_\ell(O_2).
\]
We proceed by induction over a topological ordering of the sink-relevant subgraph. The base case holds for all leaves in $L_S$. For an internal node $v$, all predecessor completions are equal under the induction hypothesis. Because activation guards and operator interpretation are deterministic functions of those typed predecessors, $C_v(O_1)=C_v(O_2)$. The argument therefore reaches every declared sink, giving
\[
\forall s\in S,\qquad C_s(O_1)=C_s(O_2).
\]
The reconstruction contract and determinism then imply
\[
\operatorname{FactSem}(A_{\mathrm{out}}(O_1))
=
\operatorname{FactSem}(A_{\mathrm{out}}(O_2)),
\]
which proves Proposition~\ref{prop:purpose-noninterference}.

The equality is semantic rather than byte-level: framework-specific control fields, serialization details, or non-factual provenance may differ without changing the factual reusable state. The property therefore captures the intended privacy boundary---changes to observation semantics that do not alter the purpose-relevant leaf results cannot alter released factual semantics.

\section{Evaluation Protocol}
\label{app:evaluation}

\subsection{Main metrics and evaluator}
\label{app:metrics}

We report Privacy, Task Success Rate, and Privacy-Compliant Success (PCS) per case:
\[
\operatorname{PCS}(i)
=
\operatorname{Privacy}(i)
\land
\operatorname{TaskSuccess}(i).
\]
Privacy checks whether downstream-visible reusable state avoids explicitly disallowed content and semantically unauthorized payloads. Task Success Rate checks whether all task-required results remain available in downstream-visible semantic content or target-bound typed outputs. PCS requires both conditions on the same execution. The same deterministic evaluator is applied post hoc to the fixed outputs of all methods, without output modification.

\subsection{PrivacyLens external transfer}
\label{app:privacylens-eval}

PrivacyLens contains 493 privacy-sensitive action trajectories \citep{shao2024privacylens}. We use the benchmark's action-level leakage and helpfulness judges with a fixed local Mistral-7B-Instruct-v0.2 evaluation model. The Purpose IR frontend is Qwen3.5-0.8B, and the Purpose runtime does not receive benchmark sensitive labels or final-action labels. The downstream action model receives only the authorized sink artifact.

We report Privacy Pass, Helpfulness $\geq 2$, Mean Helpfulness, and Safe Action Rate. Safe Action Rate is computed per trajectory as
\[
\operatorname{SafeAction}(i)
=
\operatorname{PrivacyPass}(i)
\land
[\operatorname{Helpfulness}(i)\geq 2].
\]
\subsection{Baseline boundary}
\label{app:baselines}

\begin{table}[t]
\centering
\caption{Operational boundary used for the compared methods.}
\label{tab:baseline-boundary}
\small
\begin{adjustbox}{max width=\linewidth}
\begin{tabular}{lp{0.72\linewidth}}
\toprule
Method & Evaluation boundary \\
\midrule
No Guard & Uses the fixed workflow artifact without privacy filtering. \\
Presidio & Applies the official analyzer/anonymizer to artifact content. \\
IBM Contextual Privacy LLM & Uses the contextual-privacy component with its fixed model configuration. \\
Clean Task-text Guard & Uses Qwen3.5-0.8B with the natural-language task as direct filtering context, without Purpose IR authority. \\
Fides constrained inspection & Instantiates a Fides-based IFC/constrained-inspection path with Qwen3.5-0.8B semantic inspection. \\
AgentSpec manual privacy rule & Applies an independently specified AgentSpec-style static privacy-rule interpreter. \\
Purpose IR & Uses the task-only semantic frontend, deterministic Purpose compiler, typed evidence binding, topological execution, sink validation, and release reconstruction. \\
\bottomrule
\end{tabular}
\end{adjustbox}
\end{table}

All methods operate at the same fixed workflow-artifact boundary used by Purpose IR. Appendix~\ref{app:repro} records model roles and execution settings.

\section{Additional Results}
\label{app:results}

\subsection{Main aggregate results}
\label{app:aggregate}

\begin{table}[t]
\centering
\caption{Aggregate results over the 1,200-case main workflow-state benchmark.}
\label{tab:appendix-main-results}
\small
\begin{adjustbox}{max width=\linewidth}
\begin{tabular}{lrrr}
\toprule
Method & Privacy & Task Success Rate & PCS \\
\midrule
No Guard & 296 (24.7) & 966 (80.5) & 287 (23.9) \\
Presidio & 312 (26.0) & 852 (71.0) & 249 (20.8) \\
IBM Contextual Privacy LLM & 588 (49.0) & 769 (64.1) & 387 (32.3) \\
Clean Task-text Guard & 1155 (96.2) & 318 (26.5) & 304 (25.3) \\
Fides constrained inspection & 1189 (99.1) & 671 (55.9) & 670 (55.8) \\
AgentSpec manual privacy rule & 1052 (87.7) & 300 (25.0) & 195 (16.2) \\
\textbf{Purpose IR} & \textbf{1195 (99.6)} & \textbf{1000 (83.3)} & \textbf{1000 (83.3)} \\
\bottomrule
\end{tabular}
\end{adjustbox}
\end{table}

Purpose IR reaches 83.3\% PCS with 99.6\% Privacy and 83.3\% Task Success Rate. Fides is the strongest high-privacy baseline at 99.1\% Privacy and 55.8\% PCS, giving Purpose IR a 27.5-point PCS advantage at a similar privacy level.

\begin{table}[t]
\centering
\caption{Per-workflow Purpose IR results on the main 1,200-case benchmark. Counts are shown with percentages in parentheses. Valid IR reports successful static validation of the compiled Purpose program.}
\label{tab:per-workflow-results}
\small
\begin{adjustbox}{max width=\linewidth}
\begin{tabular}{lrrrr}
\toprule
Workflow & Privacy & Task Success & PCS & Valid IR \\
\midrule
LangGraph & 398 (99.5) & 342 (85.5) & 342 (85.5) & 400 (100.0) \\
CrewAI & 398 (99.5) & 319 (79.8) & 319 (79.8) & 400 (100.0) \\
MobileRun & 399 (99.8) & 339 (84.8) & 339 (84.8) & 400 (100.0) \\
\midrule
\textbf{Aggregate} & \textbf{1195 (99.6)} & \textbf{1000 (83.3)} & \textbf{1000 (83.3)} & \textbf{1200 (100.0)} \\
\bottomrule
\end{tabular}
\end{adjustbox}
\end{table}

\subsection{Runtime mechanism ablations}
\label{app:mechanism-details}

We isolate the contributions of recovery binding and sink-constrained reconstruction while preserving the compiled programs and common evaluator.

\begin{table}[t]
\centering
\caption{Paired runtime ablations on the main benchmark. Counts are out of 1,200 cases; percentages are shown in parentheses.}
\label{tab:clean-mechanism-ablations}
\small
\begin{adjustbox}{max width=\linewidth}
\begin{tabular}{lrrr}
\toprule
Variant & Privacy & Task Success Rate & PCS \\
\midrule
Full Purpose IR & 1195 (99.6) & 1000 (83.3) & 1000 (83.3) \\
w/o recovery binding & 1198 (99.8) & 162 (13.5) & 162 (13.5) \\
w/o sink-constrained reconstruction & 419 (34.9) & 854 (71.2) & 290 (24.2) \\
\bottomrule
\end{tabular}
\end{adjustbox}
\end{table}

Removing recovery binding isolates the contribution of requirement-directed evidence recovery, while removing sink-constrained reconstruction tests whether restricting reusable state to validated sink outputs is necessary for privacy.

\section{Validation of IR Semantics}
\label{app:language-checks}

\subsection{Static validation}
\label{app:validation-checks}

Static validation enforces graph well-formedness, typed dependency constraints, raw-access restrictions, sink validity, and release ordering. Programs that violate these constraints are rejected before execution, supporting the structural assumptions used by the formal analysis.

\subsection{Cardinality semantics}
\label{app:cardinality-checks}

The main benchmark contains three cases in which a scalar requirement encounters multiple distinct typed candidates. Full Purpose IR enforces \texttt{cardinality=one} and withholds ambiguous scalar release. Removing the cardinality constraint changes the aggregate result to 1185/1200 Privacy, 1014/1200 Task Success Rate, and 1005/1200 PCS. The comparison captures the intended trade-off: releasing an ambiguous multi-value result can recover a task outcome while violating the declared scalar contract.

\subsection{Flat state schema}
\label{app:flat-schema}

The main benchmark's canonical programs do not require \texttt{COMPOSE}, multi-leaf sinks, or graph depth of at least two. Replacing the graph representation with a flat state schema therefore produces the same aggregate Privacy, Task Success Rate, and PCS as Full Purpose IR on this distribution. We evaluate compositional language semantics separately below.

\subsection{Controlled compositional suite}
\label{app:compositional-frontend}

The suite contains 60 semantic scenarios, each instantiated with five paraphrases, yielding 300 task instances. It covers argument binding, record composition, conditional routing, explicit grouping, and multi-stage dependency. Semantic-contract scoring is invariant to node identifiers and serialization order.

Of the 300 instances, 270 use two evidence leaves with depth-one dependency structure, while 30 exercise grouping or two-stage dependencies with three leaves and depth two. The 0.8B semantic-proposal frontend satisfies 182/270 depth-one instances and 12/30 deeper instances; the 27B frontend reaches 266/270 and 22/30, respectively.

\begin{table}[t]
\centering
\caption{Frontend/compiler results on the controlled compositional suite.}
\label{tab:compositional-frontend}
\small
\begin{adjustbox}{max width=\linewidth}
\begin{tabular}{lrrr}
\toprule
Frontend / compilation path & Valid IR & Semantic contract & Privacy \\
\midrule
Minimal full-IR, 0.8B & 206/300 & 25/300 (8.3) & 300/300 \\
Balanced full-IR, 0.8B & 234/300 & 139/300 (46.3) & 300/300 \\
Semantic proposal + lowering, 0.8B & 293/300 & 194/300 (64.7) & 300/300 \\
Semantic proposal + lowering, 27B & 298/300 & \textbf{288/300 (96.0)} & 300/300 \\
Gold IR & 300/300 & 300/300 (100.0) & 300/300 \\
\bottomrule
\end{tabular}
\end{adjustbox}
\end{table}

\subsection{PrivacyLens external transfer}
\label{app:privacylens-results}

\begin{table}[t]
\centering
\caption{External transfer on 493 PrivacyLens trajectories. Safe Action requires both Privacy Pass and Helpfulness $\geq 2$ on the same trajectory.}
\label{tab:appendix-privacylens}
\small
\begin{adjustbox}{max width=\linewidth}
\begin{tabular}{lrrrr}
\toprule
Method & Privacy Pass & Help. $\geq2$ & Safe Action & Mean Help. \\
\midrule
Full trajectory & 308 (62.5) & 487 (98.8) & 304 (61.7) & 2.375 \\
Task-text reduction & 365 (74.0) & 455 (92.3) & 330 (66.9) & 2.191 \\
Presidio & 392 (79.5) & 400 (81.1) & 307 (62.3) & 1.870 \\
AgentSpec manual privacy rule & 449 (91.1) & 321 (65.1) & 277 (56.2) & 1.499 \\
Fides constrained inspection & 473 (95.9) & 184 (37.3) & 168 (34.1) & 0.826 \\
\textbf{Purpose IR} & 427 (86.6) & 444 (90.1) & \textbf{381 (77.3)} & 2.081 \\
\bottomrule
\end{tabular}
\end{adjustbox}
\end{table}

\subsection{Representative semantic-boundary examples}
\label{app:examples}

\begin{table}[t]
\centering
\caption{Representative observation-to-state reductions.}
\label{tab:examples}
\small
\begin{adjustbox}{max width=\linewidth}
\begin{tabular}{p{0.23\linewidth}p{0.43\linewidth}p{0.25\linewidth}}
\toprule
User task & Rich observation may contain & Authorized result \\
\midrule
Check whether I have a meeting reminder. & Sender, time, location, phone number, full notification body. & \texttt{EXISTS(meeting\_reminder)} $\rightarrow$ Boolean. \\
\addlinespace
Has my package been delivered? & Retailer, delivery address, timestamp, notification prose. & Typed delivery-status result. \\
\addlinespace
Retrieve the requested meeting link. & URL, phone number, access code, surrounding message text. & Unique validated \texttt{meeting\_link} scalar. \\
\bottomrule
\end{tabular}
\end{adjustbox}
\end{table}

\section{Implementation and Evaluation Details}
\label{app:repro}

\subsection{Model roles}
\label{app:model-roles}

\begin{table}[t]
\centering
\caption{Model and deterministic component roles.}
\label{tab:model-roles}
\small
\begin{adjustbox}{max width=\linewidth}
\begin{tabular}{ll}
\toprule
Role & Model or execution mode \\
\midrule
Base-case upstream workflow generation & Fixed workflow artifacts generated by the 35B model \\
Controlled stress cases & Deterministic workflow-native state \\
Task semantic frontend & Qwen3.5-0.8B, task-only input \\
Purpose compilation & Deterministic lowering and graph validation \\
Evidence binding & Deterministic workflow adapters \\
Purpose interpreter & Deterministic \\
Main evaluator & Deterministic \\
PrivacyLens judge & Mistral-7B-Instruct-v0.2, evaluation only \\
Fides constrained inspection & Qwen3.5-0.8B semantic inspection in independent baseline \\
IBM baseline & \texttt{llama3.2:3b} in fixed contextual-privacy component \\
\bottomrule
\end{tabular}
\end{adjustbox}
\end{table}

The neural frontend proposes task semantics before compilation. Binding, typed execution, sink validation, and release reconstruction are deterministic and do not use task text, benchmark labels, or case identifiers as independent authorization sources.

\subsection{Implementation scope}
\label{app:scope}

The implemented system realizes the bounded Purpose IR described in this paper: task-only semantic proposal, deterministic graph compilation and validation, workflow-specific typed evidence adapters, topological interpretation, structured sinks, and reconstruction-based release. The language represents authorized semantic dataflow rather than general agent control flow; tool planning, loops, retries, and arbitrary execution strategy remain in the host agent.

\end{document}

%% file: math_commands.tex
%%%%% NEW MATH DEFINITIONS %%%%%

\usepackage{amsmath,amsfonts,bm}

% Mark sections of captions for referring to divisions of figures

% Highlight a newly defined term

% Figure reference, lower-case.

% Figure reference, capital. For start of sentence

% Section reference, lower-case.

% Section reference, capital.

% Reference to two sections.

% Reference to three sections.

% Reference to an equation, lower-case.
\def\eqref#1{equation~\ref{#1}}
% Reference to an equation, upper case

% A raw reference to an equation---avoid using if possible

% Reference to a chapter, lower-case.

% Reference to an equation, upper case.

% Reference to a range of chapters

% Reference to an algorithm, lower-case.

% Reference to an algorithm, upper case.

% Reference to a part, lower case

% Reference to a part, upper case

\def\1{\bm{1}}

% Random variables

% rm is already a command, just don't name any random variables m

% Random vectors

% Elements of random vectors

% Random matrices

% Elements of random matrices

% Vectors

% Elements of vectors

% Matrix

% Tensor
\DeclareMathAlphabet{\mathsfit}{\encodingdefault}{\sfdefault}{m}{sl}
\SetMathAlphabet{\mathsfit}{bold}{\encodingdefault}{\sfdefault}{bx}{n}

% Graph

% Sets

% Don't use a set called E, because this would be the same as our symbol
% for expectation.

% Entries of a matrix

% entries of a tensor
% Same font as tensor, without \bm wrapper

% The true underlying data generating distribution

% The empirical distribution defined by the training set

% The model distribution

% Stochastic autoencoder distributions

 % Laplace distribution

% Wolfram Mathworld says $L^2$ is for function spaces and $\ell^2$ is for vectors
% But then they seem to use $L^2$ for vectors throughout the site, and so does
% wikipedia.

 % See usage in notation.tex. Chosen to match Daphne's book.